\documentclass[3p]{elsarticle}

\usepackage{amssymb}
\usepackage{amsmath}
\usepackage{amscd}

\def\Journal#1#2#3#4{{#1} {\bf #2}, (#3) #4}

\def\PMB{\em Phys. Med. Biol.}
\def\MP{\em Med. Phys.}
\def\ARI{\em Appl. Radiat. Isot.}

\def\NIMA{{\em Nucl. Instrum. Methods} \bf A}
\def\NIMB{{\em Nucl. Instrum. Methods} \bf B}
\def\JRP{\em J. Radiol. Prot.}
\def\Radiology{\em Radiology}

\begin{document}

\begin{frontmatter}

\title{Monte Carlo study of a 3D Compton imaging device with GEANT4}

\author[fi]{M. Lenti}
\ead{lenti@fi.infn.it}
\author[ub,fi]{M.~Veltri\corref{cor1}}
\ead{michele.veltri@uniurb.it}
\cortext[cor1]{Corresponding author}
\address[fi]{Sezione dell'INFN di Firenze, via G. Sansone 1, I-50019 Sesto F.(FI), Italy }
\address[ub]{Dipartimento di Matematica, Fisica e Informatica, Universit\`a di Urbino, 
via S. Chiara 27, I-61029 Urbino, Italy }

\begin{abstract}
In this paper we investigate, with a detailed Monte-Carlo simulation 
based on Geant4, the novel approach  \cite{lenti} to 3D imaging with 
photon scattering. 
A monochromatic and well collimated gamma beam is used to
illuminate the object to be imaged and the photons Compton scattered
are detected by means of a surrounding germanium strip detector.
The impact position and the energy of the photons are measured
with high precision and the scattering position along the beam axis 
is calculated.
We study as an application of this technique the case of brain 
imaging but the results can be applied as well to situations where a 
lighter object, with localized variations of density, is embedded 
in a denser container.
We report here the attainable sensitivity in the detection of
density variations as a function of the beam energy, the depth
inside the object and size and density of the inclusions.
Using a 600~keV gamma beam, for an inclusion with a density 
increase of 30\% with respect to the sorrounding tissue and 
thickness along the beam of 5~mm, we obtain at midbrain position 
a resolution of about 2~mm and a contrast of 12\%.
In addition the simulation indicates that for the same gamma 
beam energy a complete brain scan would result in an effective 
dose of about 1~mSv.

\end{abstract}


\end{frontmatter}

\section{Introduction}
\label{introduction} 
For long time Compton scattering has been widely employed
as an imaging tool. A photon in the energy range 0.1$\div$2.0~MeV
during its passage through matter, will dominantly Compton scatter against
the electrons of the material with an intensity proportional to electron
(and mass) density. If the energy of the incident and scattered photons are 
known, it is possible to locate the scatter volume using the well known 
energy--angle relationship. Photon scattering can therefore readily provide 
3D density maps of extended objects of unknown composition. 
This is a clear advantage of scattering compared to other photon 
imaging techniques, like the well established x--ray applications, which
make use of photons transmitted through the material to be imaged. 
In this case the integrated density of the object along the path traversed 
by the photon beam is provided. Several profiles from different directions 
need to be collected and complex algorithms are required in order 
to reconstruct the 3D density  distribution.\\
In the last fifty years Compton scattering has found a broad range of
applications in medicine, as imaging tool \cite{lale}--\cite{elkhettabi} 
or to measure the human tissue density \cite{albahari}. 
Compton scattering is also widely used in the field of non destructive 
testing (NDT) for quality control to detect defects like cracks, inclusions 
and voids inside industrial manufactures \cite{harding}--\cite{ho}.
The possibility of using backscattered radiation has made Compton 
scattering very attractive also in those situations where the access to 
the object to be examinated is restricted to one side, like inspection
of airframes \cite{evans} and soil density determination \cite{balogun}.\\
More recently attention has been focused on reconstructing the photon 
arrival direction (for instance from implanted nuclides) exploiting 
Compton scattering in a position sensitive detector 
\cite{wulf}--\cite{marel}: this  method allows a conical 
region of possible photon origin to be defined, and with several 
measurements the true origin can be inferred by intersection of 
different cones.\\
In spite of its attractive features, photon scattering has not
become the dominant approach in the imaging field  \cite{hussein}.
There are two major limitations of this technique:
the background induced by events multiply scattered inside
the object and the attenuation of both incident and scattered
photons. Many attempts to overcome these limitations have been pursued.
We recall the use of the information of the transmitted beam to determine 
the attenuation coefficient, or the use of collimators to define a small 
region of interest in the examinated object in order to reduce the 
multiple scatter component. 
Also the Monte Carlo approach has proven to be able to provide 
corrections for this source of background \cite{speller,mooney}.\\
In this paper we want to further explore the potential of the novel method 
to image the electron density in the human body proposed in~\cite{lenti}
by means of a detailed Monte Carlo (MC) simulation using GEANT4~\cite{geant4}
with a realistic set--up. The application presented here refers to brain 
imaging but the results can be applied as well to situations where a 
lighter object, with localized variations of density, is embedded in a 
denser container.\\
This paper is organized as follows: in Section \ref{sec:compton} we 
will briefly recall the main characteristics of Compton scattering,
in Section \ref{sec:exptec} the experimental technique will be described
and in Section \ref{sec:mcsim} details about the simulation will be given.
In Section \ref{sec:results} we will describe the results found and the 
expected performance of the proposed device, in Section \ref{sec:dose}
we will give an evaluation of the expected dose, and finally in Section 
\ref{sec:conclusions} we will give our conclusions.

\section{Compton Scattering}
\label{sec:compton} 
In the Compton effect a photon of energy $E$
interacts with an atomic electron and it is scattered,
with a lower energy $E^{\prime}$, at an angle
$\theta$  with respect to the incoming direction. 
In case of scattering by free electrons at rest the conservation of energy 
and momentum implies the well--known Compton formula which relates 
initial and final photon energies to the scattering angle:
\begin{equation}
\label{eq:compton}
   \cos\theta = 1 + \frac{m_e}{E} - \frac{m_e}{E^{\prime}},
\end{equation}
where $m_e= 511$ keV is the electron mass (natural units with
speed of light $c=1$ are used). The kinetic energy $T$ of the struck
electron is $T=E-E^\prime$ and the maximum is:
\begin{equation}
\label{eq:tmax}
   T_{max} = \frac{E}{1+\frac{m_e}{2E}},
\end{equation}
The probability of interaction is given by the Klein-Nishina formula 
\cite{bjorken}:
\begin{equation}
\label{eq:klein}
   \frac{d\sigma}{d\Omega} = 
\frac{r_e^2}{2}\left(\frac{E^\prime}{E}\right)^2\left[\frac{E^\prime}{E}+\frac{E}{E^\prime}-\sin^2\theta\right]
\end{equation}
where $r_e= 2.818\times 10^{-15}$ m is the electron classical
radius. Eq.~\ref{eq:compton} and \ref{eq:klein} are however only approximations.
In reality electrons are not free and transitions of bound electrons 
are allowed only if the energy transfer is larger than the ionization 
energy.  Moreover electrons are not at rest but move with a certain 
momentum distribution. This produces  the so called 
''Doppler broadening'' i.e. a smearing with a long, non gaussian tail, 
of the scattering angle. Both the binding effects and the Doppler broadening
are however significant at photon energies less than 100 keV, well below
energies used in the present study.

\section{Experimental technique}
\label{sec:exptec}
The proposed technique \cite{lenti} makes use of a well collimated 
and monochromatic gamma-emitting source which illuminates the 
part of the body/object to be imaged.
An x-y-z orthogonal coordinate system is introduced such that the $z$ 
axis is along the beam direction.  Photons Compton scattered 
inside the body/object are recorded by means of a surrounding germanium
strip detector in the position 
$(x^\prime,y^\prime,z^\prime)$ and its energy $E^\prime$ is measured.
\begin{figure}[t]
 \begin{center}
  \vskip -7.5cm
  \includegraphics[width=14cm]{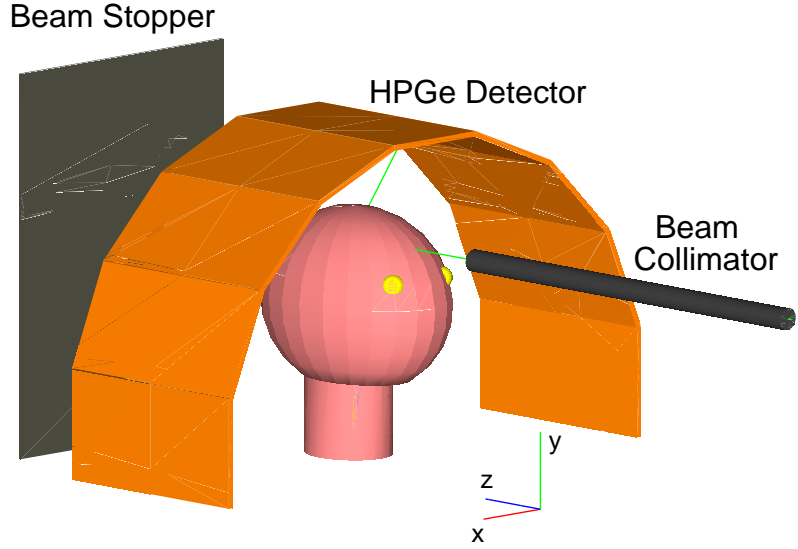}
    \vskip -2.0cm
    \caption{A collimated gamma beam is directed towards an head phantom and 
    the scattered photons are observed with a germanium strip detector. The
    beam is along the z axis. A lead plate in the rear part of the device 
    acts as a beam stopper.}
   \label{fig:set-up}
 \end{center}
\end{figure}
Using Eq.~\ref{eq:compton} the  angle $\theta$ is calculated
and the scattering position $z$ is reconstructed by the formula: 
\begin{equation} 
  \label{eq:z}
   z = z^{\prime} - \cot\theta\sqrt{{\left(x^{\prime}-x_{beam} \right)}^2 + {\left(y^{\prime}-y_{beam} \right)}^2}
\end{equation}
The $z$ distribution so obtained provides, directly and without any inversion
algorithm, the density distribution of
the material crossed by the beam i.e. a small cylinder with
transverse dimensions given by the beam size and length given by
the object length. A full imaging is obtained by raster scanning 
the surface of the object.
Typical scattering applications make use of collimators in front of the 
detector to precisely define the scatter volume (voxel), given by 
the intersection of the incident and detected photon. 
The method proposed here is collimator free. 
This aspect provides larger counting rates and therefore lower dose 
absorbed by the patient.

\section{Simulation}
\label{sec:mcsim} 
The study was developed in the  GEANT4~\cite{geant4} framework. 
The set--up used for the simulation (see fig.~\ref{fig:set-up}) consists 
of a monochromatic gamma source placed at 555~mm from the head
center, which is positioned in $z=0$, and at  5~mm distance from a collimator. 
This is a lead tube 400~mm long and of inner and outer radius 1~mm 
and 10~mm, respectively.
The obtained gamma beam is directed onto the head volume.
The modelling of the human head is based on the ''human-phantom'' 
example provided with the GEANT4 package, the anthropomorphic MIRD 
\cite{mird} phantom was used.
The shape of the head is a cut ellipsoid and consists of three
nested structures. 
Externally we find  a skin layer made by soft tissue, then the skull 
volume made by bone tissue with density $\rho_S$=1.486~g/cm$^3$.
The skull contains the brain volume made by soft tissue with density
 $\rho_B$=1.040~g/cm$^3$ (the same density was used for the skin). 
To test the capability of the apparatus to detect inclusions of
different density, we positioned inside the brain volume and on the
beam axis, four small target volumes of 2x2~mm$^2$ in the transverse 
beam direction and 5~mm in the longitudinal one. 
These targets are made by soft tissue but a greater value for the
density of $\rho_T$=1.35~g/cm$^3$ was used. This value was chosen in 
accordance with magnetic resonance measurements of the density of 
brain pathological tissues \cite{tumor}.
The targets (numbered 0--3) were placed at the $z$ positions 
-45; -15, 0 ,15 mm along the beam line.\\
Photon detection is achieved by means of a segmented HPGe detector. 
\begin{figure}
 \begin{center}
  \includegraphics[width=10cm]{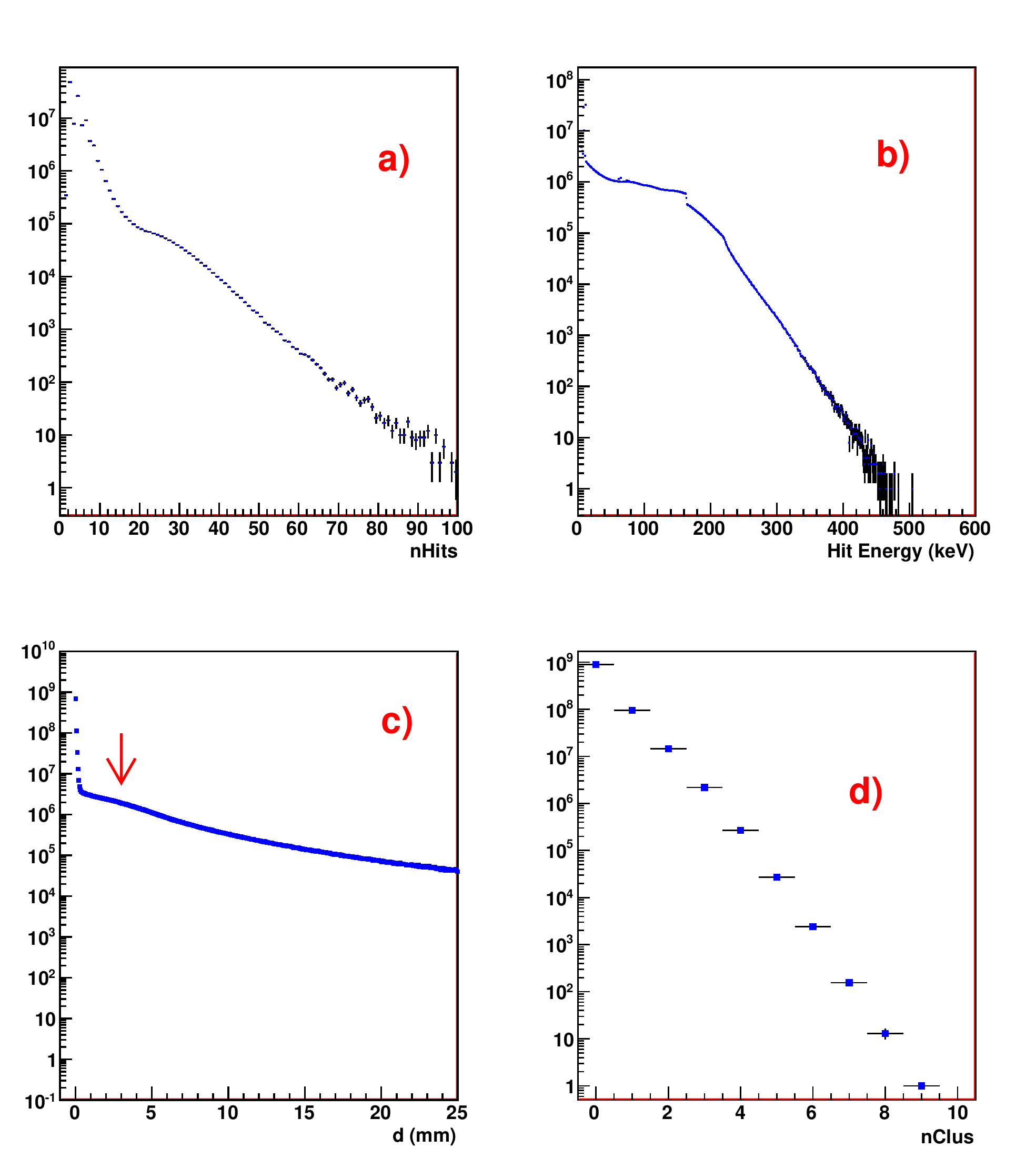}
   \caption{Distribution of the number of hits per event in the HPGe detectors (a), 
    hit energy (b), distance between hits (c),  
    number of reconstructed clusters per event (d).
    The arrow indicates the maximum distance for clustering together hits.
    A beam energy of 600 keV was used, $10^9$ events were simulated. }
    \label{fig:hpge_cluster}
 \end{center}
\end{figure}
Two half--rings of such devices surround the head volume. 
The sensors are parallelepiped 5 mm thick, 100 mm wide and 100 mm high 
placed according to an hexadecagon shape to cover the largest possible 
azimuthal region. A 5~mm thick lead plate placed in the rear part of 
the head acts as a beam stopper. Each simulation consisted of $10^9$ 
events. Ten different gamma energies were simulated from 100 to 1000 
keV in steps of 100 keV.
Interactions of photons and electrons were simulated using the GEANT4 
implementations of physics models developed for the PENELOPE code 
\cite{penelope}. This code has been specifically developed for $\gamma$ and
$e^{\pm}$ transport in matter and great care was given to the description 
of the low energy processes including binding effects and Doppler broadening.
Comparisons with exeprimental data showed that PENELOPE provides consistent
results in the energy range from few keV up to about 1~MeV \cite{sempau}.\\
Photons were tracked through the described set-up. Step by step the type 
of interaction, its position and the volumes crossed were recorded together 
with the energy deposited in the head volumes for dose calculations. 
Hits produced in the HPGe detectors were also recorded.
The energy resolution of the sensors was taken to be gaussian and was 
simulated taking the standard 2.96 eV energy to produce an electron-hole 
pair, a Fano factor of 0.05 and a constant electronics noise of 500 eV 
(see for example \cite{marel, knoll}, for an analysis on energy and
 angular resolution for CdZnTe detectors see \cite{du} ).
Hits within a distance of less than 3~mm were grouped together to reconstruct
a cluster. The sum of hits energies, after the smearing 
to account for the detector resolution, was taken to be the cluster energy. 
Its position is given by the energy weighted average of the hits position.
Fig.~\ref{fig:hpge_cluster} shows the distribution of the number of hits
per event in the HPGe detectors (a), their energy (b), their distance (c) 
and the number of reconstructed clusters for each scattered photon (d) 
for a beam energy of 600 keV.\\
The best $z$ resolution for the proposed setup is achieved at $\theta=90^\circ$. 
Applying error propagation on Eq.~\ref{eq:z} we obtain:
\begin{equation}
\label{eq:resolution}
   \sigma_z = \sigma_{z^{\prime}} \oplus 
              d~\sigma_{\theta{Doppler}} \oplus 
 d\frac{m_e}{E^{\prime 2}} \left( \frac{1}{sin \theta} + \frac{cos^{2} \theta}{sin^{3} \theta} \right) \sigma_{E^{\prime}} 
 \oplus 
 \overline{|\cot\theta|}
 \left(\frac{x^{\prime}-x_{beam}}{d}\sigma_{x^{\prime}} \oplus \frac{y^{\prime}-y_{beam}}{d}\sigma_{y^{\prime}} \right),
\end{equation}
where $\oplus$ means sum in quadrature. In the above equation
$\sigma_{z^{\prime}}$ is the resolution on the photon impact
position along the $z$ axis given by the strip granularity.
$d=\sqrt{(x^{\prime}-x_{beam})^2 + (y^{\prime}-y_{beam})^2}$ is the 
distance of the photon impact point on the detector from the beam axis,
for the events undergoing  a single Compton scattering (see 
Sec.~\ref{sec:energy}) the average value in the simulated beam energy 
range (100$\div$1000~keV) varies between 141 and 148~mm; 
$\sigma_{\theta{Doppler}}$ is the Doppler broadening 
contribution, it goes from $6^\circ$ to about $3^\circ$ depending on the 
beam energy.
Both quantities show a steep energy dependence reaching a plateau at about 500~keV.
$\sigma_{E^{\prime}}$ is the energy resolution of the germanium detector,
the term dependent on the scattering angle $\theta$ is 1 at $90^\circ$ 
and less than 2 in the range $50^{\circ}\div 125^{\circ}$. 
$\sigma_{x^{\prime}}$ and $\sigma_{y^{\prime}}$ depend on   
the germanium detector thickness and strip pitch, but their contribution 
is suppressed by a factor $\overline{|\cot\theta|}$ which is an average 
over the accepted $\theta$ range. In the simulated energy range this quantity 
goes down from 0.39 to 0.36 reaching its final value at about 500~keV.
Finally the distances in $x$ and $y$ between the beam and the detector position 
divided by $d$ are less than 1~mm, both having an average value of about 0.6~mm.\\
The error induced by the beam divergence $\sigma_{\theta{beam}}$ 
and width $\sigma_{w{beam}}$ turns out to be:
\begin{equation}
\label{eq:resolution2}
   \sigma_{z(beam)} = d~\sigma_{\theta{beam}} \oplus 
            \overline{|\cot\theta|}~\sigma_{w{beam}},
\end{equation}
where the first term is the dominant one.
Both these sources of uncertainty are determined by the collimator geometry:
$\sigma_{\theta{beam}}=2.5$~mrad and $\sigma_{w{beam}} \approx 0.5$~mm.
This formula is valid for a non attenuated beam, beam attenuation effects 
are properly taken into account by the simulation.\\
The $z$ resolution improves as $d$ decreases, i.e. the detector
must be as close as possible to the object to be imaged. 
The $z$ resolution also improves if the gamma energy increases, 
both because the energy resolution improves and because the Doppler broadening
contribution decreases. 
As it will be shown later, the number of scattered and
detected photons decreases as $E$ increases and also the
background changes. The best gamma energy is thus a compromise
among different aspects.

\section{Results of the simulation}
\label{sec:results} 
\subsection{Event types and energy distributions}
\label{sec:energy}
For the analysis we select events which have only one reconstructed 
cluster in the HPGe detector and an energy compatible with that of
Compton scattering. We classify events in four types
according to the following scheme:\\
{\it S1}: the primary $\gamma$ underwent a single Compton scattering
inside the head volume and then deposited all its energy producing
one cluster inside the HPGe.\\
{\it Sn}: in this case the $\gamma$ has undergone multiple Compton interactions
inside the head volume before releasing all its energy in the HPGe.\\
{\it Escapers}: the primary photon after one or more interactions in the head
volume gives a signal in the germanium detectors but not all of its
energy is released. The $\gamma$ or one of its daughter particles 
 escape without being detected.\\
{\it NoGamma}: for these events the signal in the HPGe was not directly
produced by the primary photon.\\
From the above description it is clear that only events of type S1
are those which carry the correct relation between energy and angle 
\begin{figure}[p]
 \begin{center}
  \includegraphics[width=13.4cm]{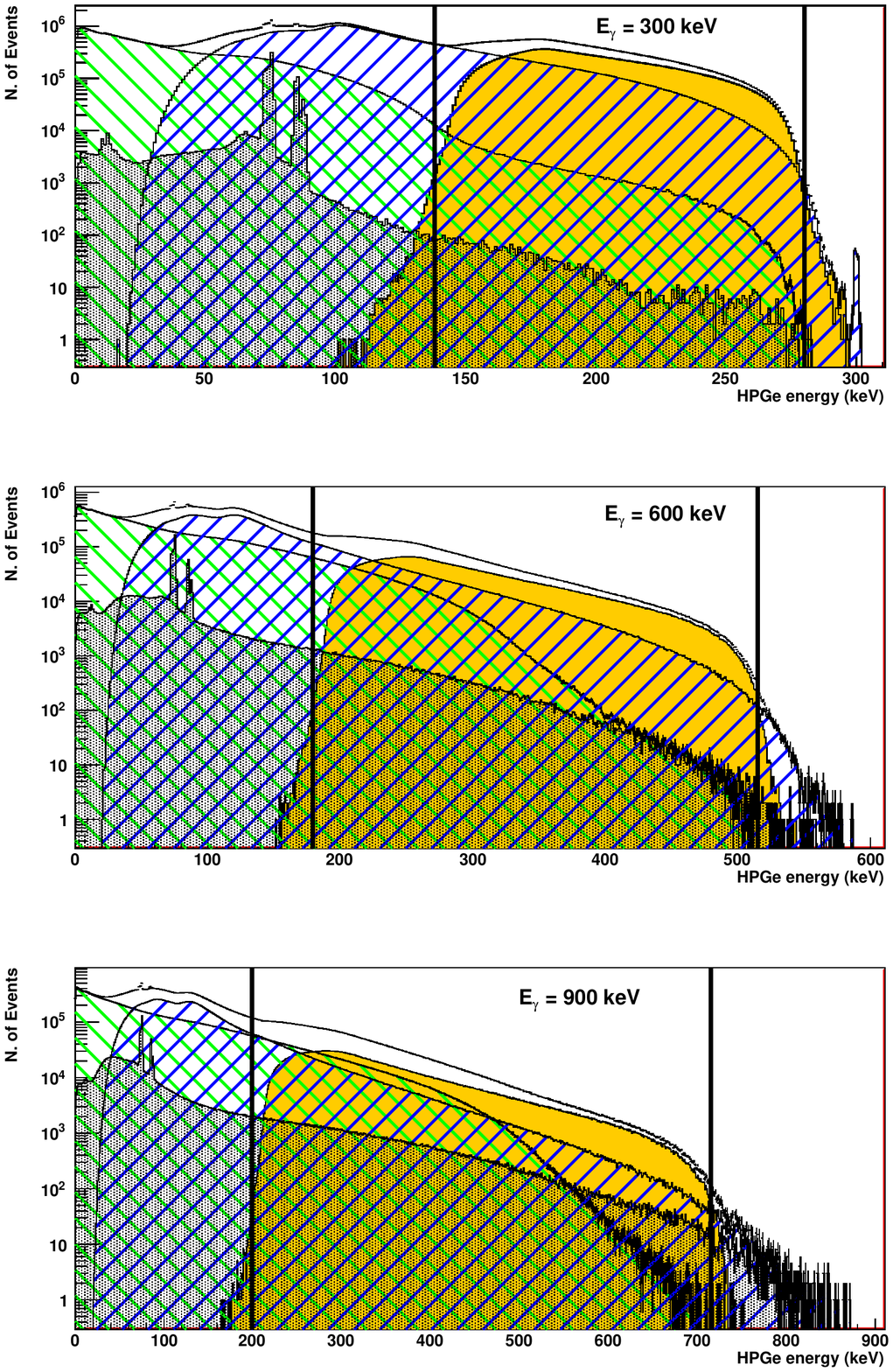}
 \caption{\it Distribution of the energy reconstructed in the HPGe 
  detector for three beam energies, from top to bottom 300, 600 and 900
  keV, respectively. The various histograms represent the different types
  of events (see text for the definition): shaded S1 type,
  45$^\circ$ hatching Sn, 145$^\circ$ hatching Escaper and heavy shaded 
  NoGamma type. The total distribution is the top histogram.
  The two vertical lines represent the energy range used for the event 
  selection. Each plot was obtained by simulating $10^9$ events.}
  \label{fig:hpge_energy}
 \end{center}
\end{figure}
and can be used to determine the $z$ position of the scattering, the 
other events constitute the background to our signal. 
Figure~\ref{fig:hpge_energy} shows the energy distribution recorded by the 
HPGe detectors for three different beam energies. The various histograms
correspond to the event type described above and the lines indicates
the energy range used for the event selection. The lower limit represents
the minimal energy required for a Compton scattering (i.e. 
$\theta=180^{\circ}$) and the upper one is the energy where the S1 
component becomes less than the Sn one. 
In the plot relative to the 300 keV beam energy is visible, at the same
energy of the beam, the peak of the coherently scattered events 
(Rayleigh peak). \\
The peak is not visible at higher energies since
the cross-section for this process decrease rapidly 
($\sigma_{Rayl} \propto E^{-2}$) at growing $\gamma$ energies.
The peak visible at the escape line of the lead at 75 keV and 85 keV in 
the distributions of NoGamma events indicate that these events are mostly 
due to electrons emitted by photoelectric effect in the beam stopper 
and backscattered onto the HPGe.\\
The percentage of the four types of events as a function of beam 
energy is given in Table~\ref{tab:events}.
\begin{table}[h]
 \begin{centering}
 \begin{tabular}{rrrrr}
  \hline
 \multicolumn{1}{c} {E$_{\gamma}$(keV)} &  \multicolumn{1}{r} {S1 (\%)}   & 
 \multicolumn{1}{r} {Sn (\%)}           &  \multicolumn{1}{r} {Escaper (\%)} & 
 \multicolumn{1}{r} {NoGamma (\%)} \\
  \hline
           100      & 71.16 & 28.37 &  0.17   & 0.30 \\
           200      & 69.91 & 29.79 &  0.30   & -- \\
           300      & 69.39 & 30.27 &  0.33   & -- \\
           400      & 69.07 & 30.32 &  0.59   & 0.02 \\
           500      & 67.92 & 29.76 &  2.20   & 0.15 \\
           600      & 65.11 & 28.28 &  6.23   & 0.37 \\
           700      & 61.38 & 26.45 &  11.47  & 0.69 \\
           800      & 57.55 & 24.51 &  16.83  & 1.11 \\
           900      & 53.87 & 22.81 &  21.72  & 1.60 \\
          1000      & 50.67 & 21.22 &  25.96  & 2.15 \\
\hline
\end{tabular}
\medskip
\caption{\it Percentage of event types as a function of the
beam energy.}
\label{tab:events}
 \end{centering}
\end{table}
\noindent We see that for beam energies up to 500 keV the event composition
is rather stable with about 70\% of events being of type S1.
For higher energies the component due to Escapers rises rapidly and at 
1~MeV the background amounts to 50\%.

\subsection{z distributions}
\label{sec:z}
\begin{figure}[h!]
 \begin{center}
  \includegraphics[width=11.3cm]{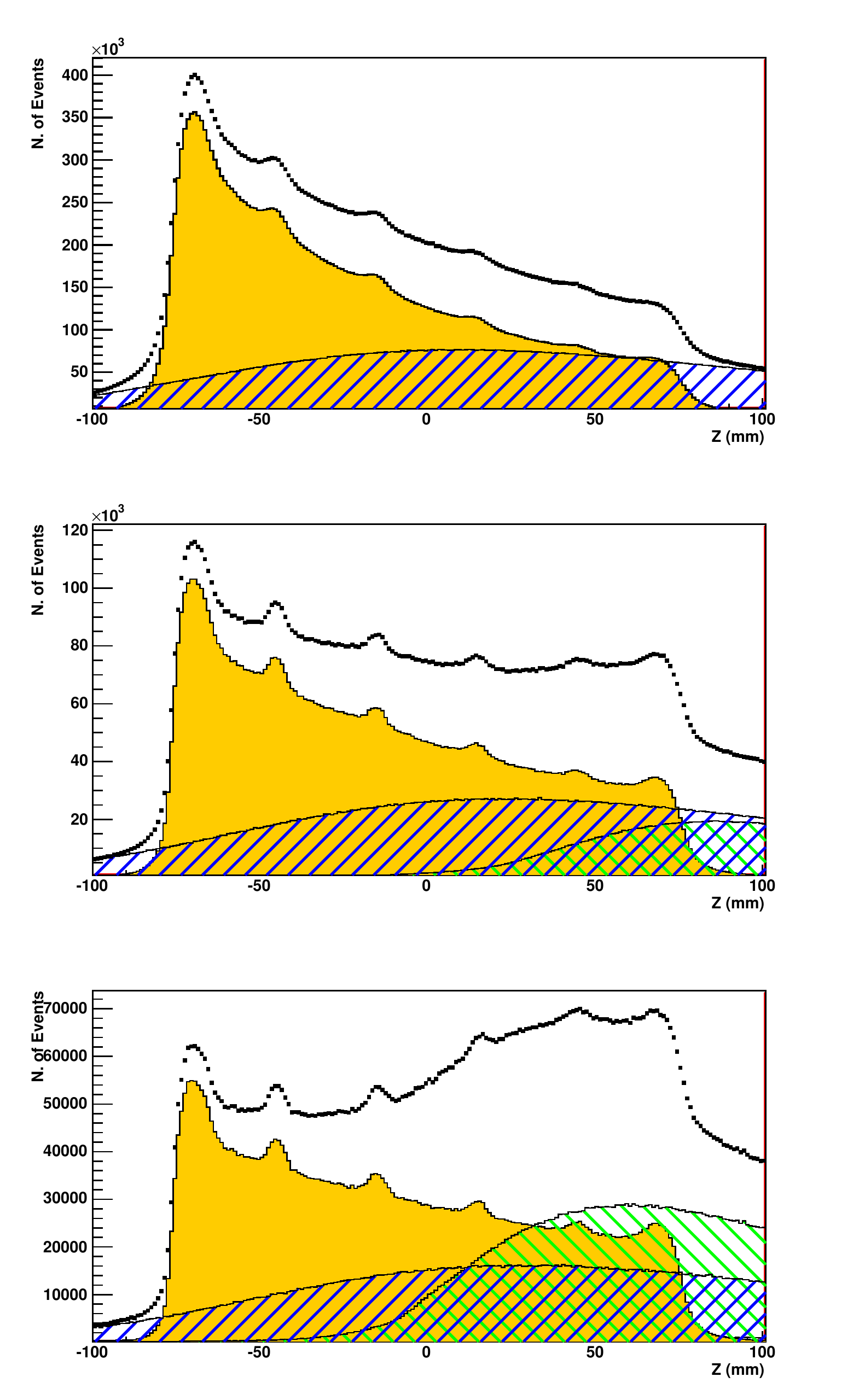}
  \caption{\it Reconstructed z position for three beam energies,
   from top to bottom 300, 600 and 900 keV, respectively. 
   The various histograms represent the  different types
   of events (see text for the definition): shaded S1 type,
   45$^\circ$ hatching Sn, 145$^\circ$ hatching Escaper type. 
   The distribution of NoGamma type events is not visible on this scale, 
   the black squares represent the total distribution. Two peaks at the 
   edges of the distribution due to the bone tissue and four due to the
   targets are visible. Each plot was obtained by simulating $10^9$ events.}
  \label{fig:zrec}
 \end{center}
\end{figure}

Figure~\ref{fig:zrec} show the distribution of the reconstructed $z$ position
for three different beam energies, from top to bottom 300, 600 and 900 keV
respectively. The various histograms refer to the different event types (see
figure caption for details).
In this picture the head volume extends from -75~mm to 75~mm, the beam 
is directed from $-z$ to $+z$.\\
Distributions are shown in the $\pm$100~mm range but they extend in 
a wider range (-200$\div$500~mm); outside the interval $\pm$80~mm only
background events contribute.
At lower $\gamma$ energies effects of attenuation are
dominant and clearly determine the shape of the distribution. 
For all beam energies the distribution of Sn events is slowly 
increasing with increasing $z$ and shows a long plateau.
The Escaper component, on the contrary, increases rapidly with beam energy 
producing a remarkable distortion on the distribution at large $z$ values.
Since the detected energy in this case is always smaller than the real one
the $\theta$ angle of the Escaper events is biased towards values greater 
than $90^{\circ}$, i.e. towards positive values of $z$.\\
The presence of localized increase of density in the head volume 
induces two effects \cite{sharaf}. An increase of the number of 
scatterings, producing a peak in the distribution, and an increase in the
attenuation of primary photons impinging on the downstream material.
This last effect is responsible for the change in the slope of the
distribution. Inclusions with lower density or voids, will 
manifest themselves as dips.
Four peaks due to the targets and two at the edges of the head volume,
due to the bone tissue, are visible in Fig.~\ref{fig:zrec}. 
They tend to disappear at large $z$ for lower beam energy. 
Higher energies on the contrary show a better penetration capability. 
We notice how the use of scattered photons
makes possible the detection of changes in the composition of the object 
to be imaged independently of beam attenuation.
With an imaging technique based on photon transmission, 
in the  case of a low density material enclosed in a dense container,
the dominant contribution to the signal would be provided by the 
container itself.
\subsection{Raw signal and resolution}
\label{sec:signal}
\begin{figure}[htb]
 \begin{center}
  \includegraphics[width=6cm]{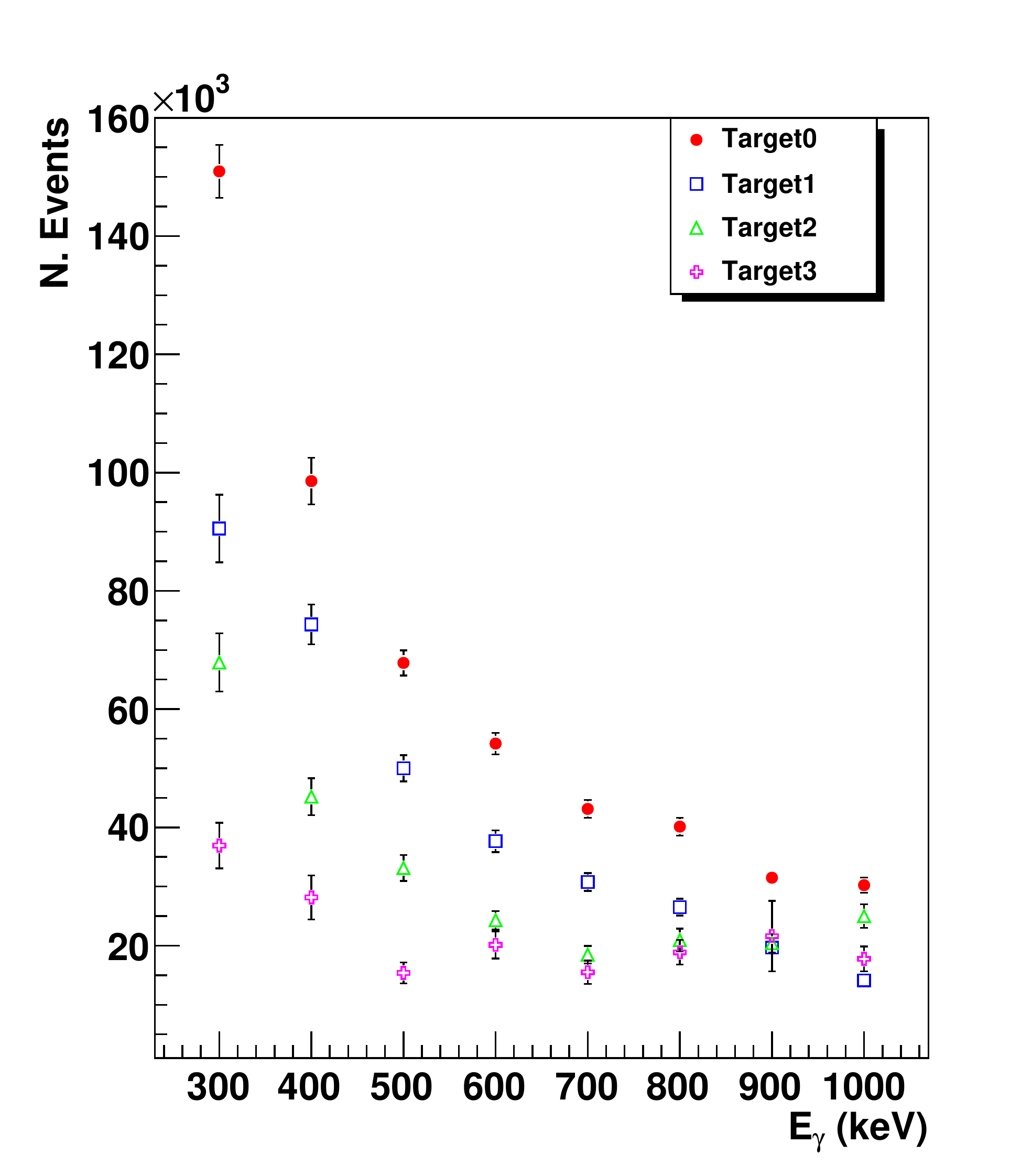}
  \caption{\it Number of events in the signal peak as a function of the 
  $\gamma$ beam energy for the four targets. For each value of beam 
  energy $10^9$ events were generated.}
  \label{fig:signalvsegamma}
 \end{center}
\end{figure}

\begin{figure}[htb]
 \begin{center}
  \includegraphics[width=7.0cm]{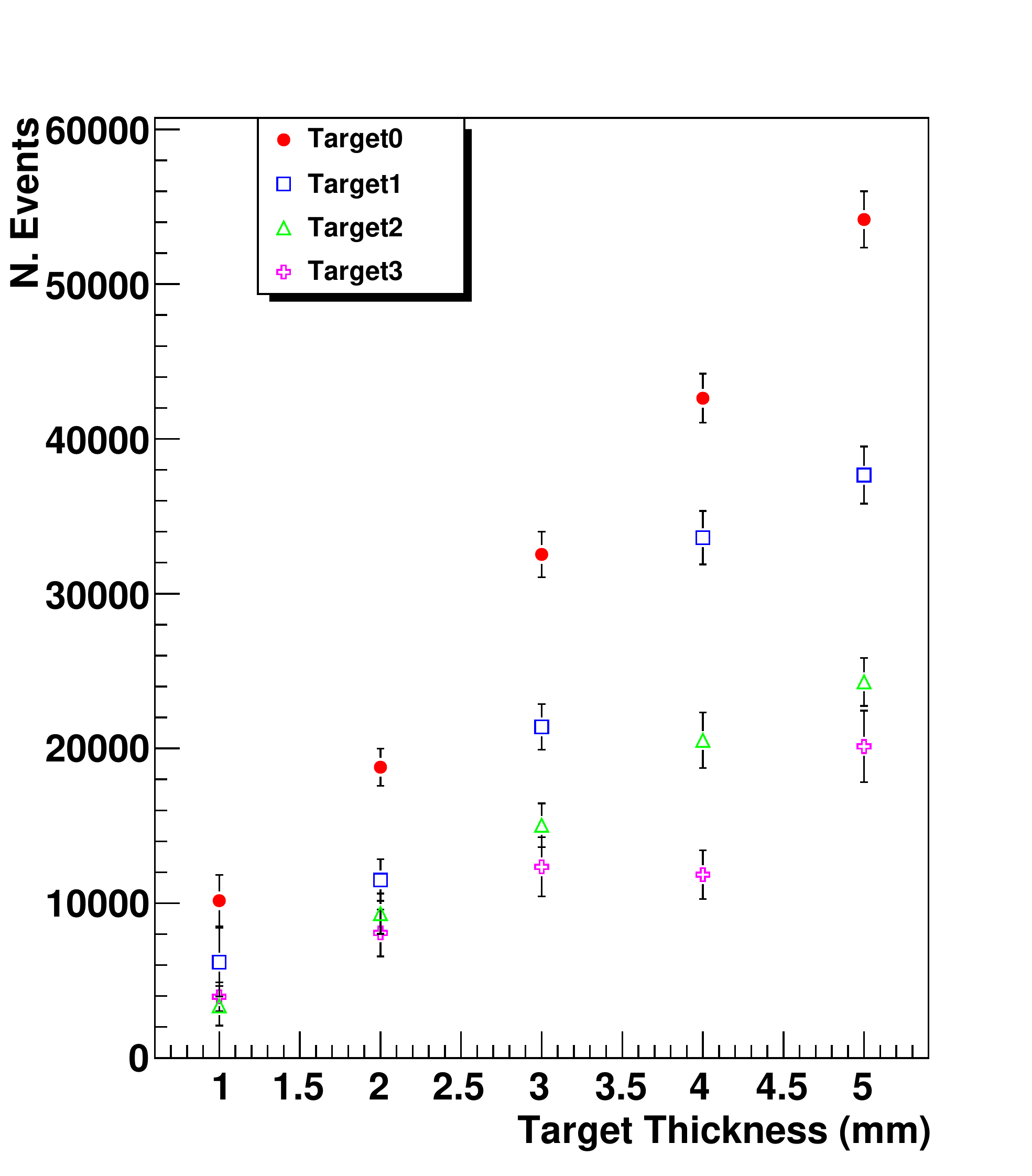}
  \caption{\it Number of events in the signal peak  as a function of the 
  target thickness for the four targets. Target 0 is the first seen by the 
  photon beam. A beam energy of 600~keV was used, for each value of target thickness 
  $10^9$ events were generated.}
  \label{fig:signalvsdz}
 \end{center}
\end{figure}

\begin{figure}[h!]
 \begin{center}
  \includegraphics[width=7.0cm]{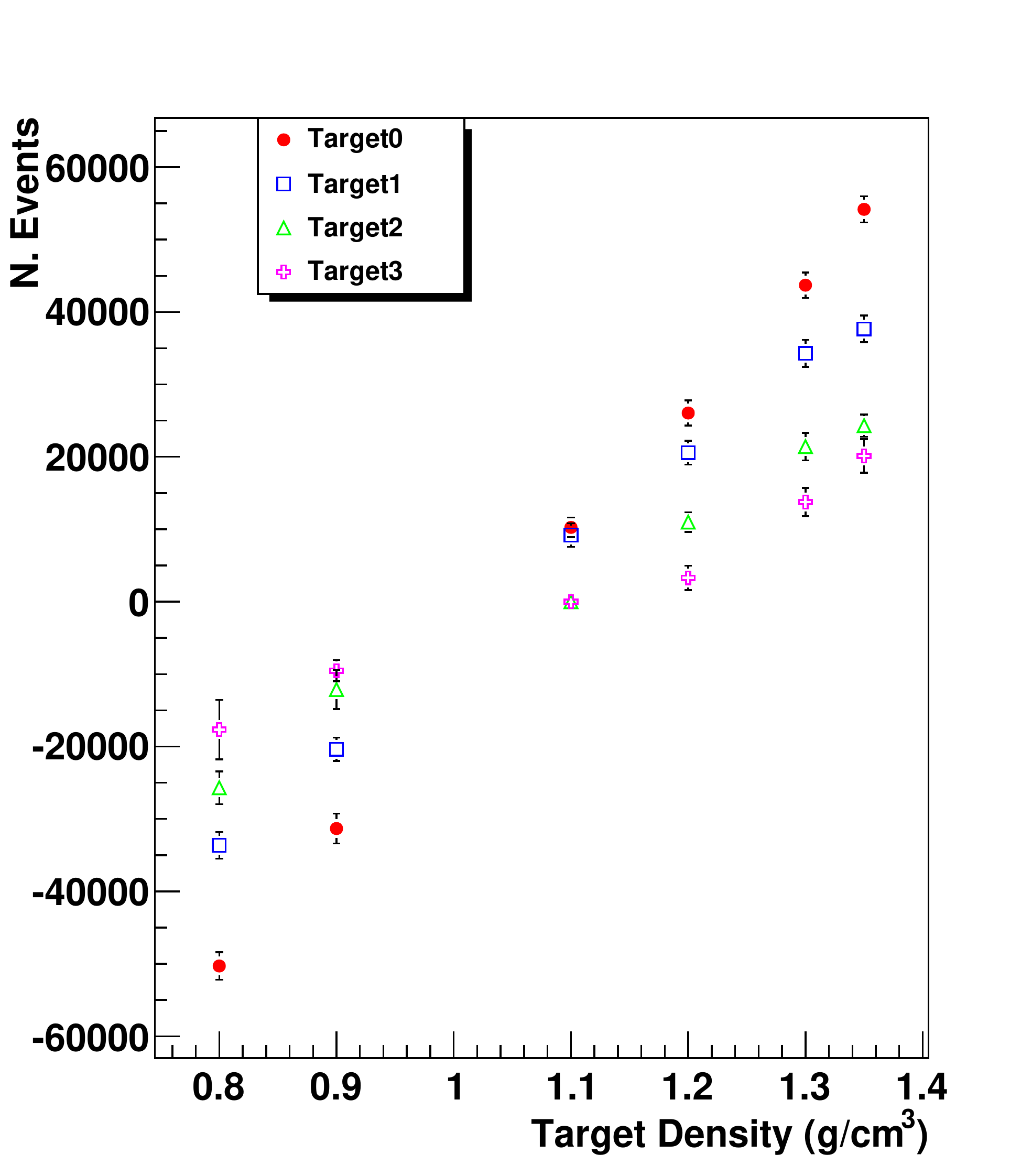}
  \caption{\it Number of events in the signal peak  as a function of the 
  target density for the four targets. The beam energy is 600 keV, for 
  each density value $10^9$ events were generated.
  The density of the material surrounding the targets is
  $\rho_B$=1.040~g/cm$^3$. Targets with smaller density will produce dips
  in the $z$ distribution. The negative number of events in the plot refers
  to the area of the reversed peak.}
  \label{fig:signalvsdensity}
 \end{center}
\end{figure}

\begin{figure}[ht!]
 \begin{center}
  \includegraphics[width=7.0cm]{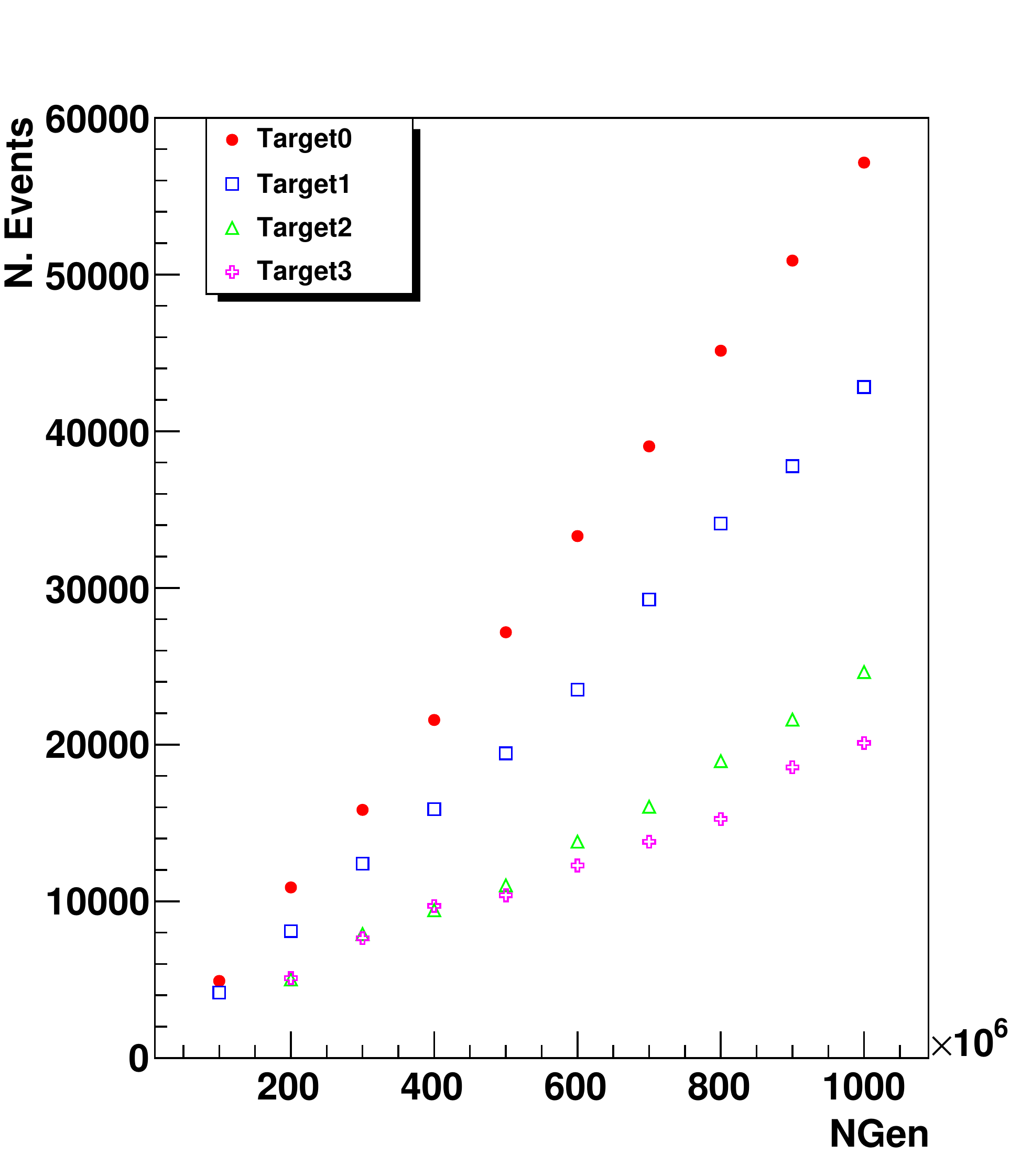}
  \caption{\it Number of events in the signal peak  as a function of 
  the number of generated events in the simulation for the four 
  targets. The beam energy is 600 keV.}
  \label{fig:signalvsngen}
 \end{center}
\end{figure}

The raw signal, i.e. not corrected for the beam attenuation, produced by 
the four targets in absence of resolution effects would be the target profile 
along $z$, a rectangle 5~mm wide with sharp edges. The resolution modifies
this profile blurring the borders. To extract the raw signal from the data we
used a model of a rectangular signal pulse convoluted with a gaussian (the
blurring function) over an exponential background (this corresponds to 
the signal of the surrounding tissue and that induced by the Sn and Escaper
events). 
The fit provides the mean position and the height of the signal together 
with the $\sigma$ of the gaussian modelling resolution effects. 
We studied the signal as a function of beam energy in the range 
300$\div$1000~keV, simulating $10^9$ events for every energy value. 
For all investigated energies the results of the fit show that the positions 
of the targets are correctly reproduced within $\approx$ 0.1~mm.
Figure~\ref{fig:signalvsegamma} shows, as a function of beam energy
and for the four targets, the number of events obtained integrating
the signal peak.
The signal is visible for all the four targets over the full energy range. 
It decreases with increasing beam energy and target $z$ position.
For the two downstream targets (2 and 3) and for energies above 600~keV 
the decrease is less pronounced due to the loss of resolution and 
background increase, which make the peak larger.
The resolution $\sigma$ as a function of beam energy is
reported in Table~\ref{tab:sigmavsegamma}.
\begin{table}[h]
 \begin{centering}
 \begin{tabular}{rrrrr}
  \hline
 \multicolumn{1}{c} {\rule{0mm}{5mm} E$_{\gamma}$(keV)} &  
 \multicolumn{1}{r} {$\sigma_{0}$ (mm)} & \multicolumn{1}{r} {$\sigma_{1}$ (mm)} & 
 \multicolumn{1}{r} {$\sigma_{2}$ (mm)} & \multicolumn{1}{r} {$\sigma_{3}$ (mm)} \\
  \hline
  300  & 2.5$\pm$0.1  &  2.5$\pm$0.2  &  3.0$\pm$0.2  &  2.9$\pm$0.3 \\
  400  & 2.2$\pm$0.1  &  2.6$\pm$0.1  &  2.6$\pm$0.2  &  2.9$\pm$0.4 \\
  500  & 1.9$\pm$0.1  &  2.1$\pm$0.1  &  2.3$\pm$0.2  &  1.6$\pm$0.4 \\
  600  & 1.9$\pm$0.1  &  1.9$\pm$0.1  &  1.8$\pm$0.2  &  2.5$\pm$0.3 \\
  700  & 1.7$\pm$0.1  &  1.8$\pm$0.1  &  1.7$\pm$0.2  &  1.9$\pm$0.3 \\
  800  & 2.0$\pm$0.1  &  1.7$\pm$0.2  &  2.1$\pm$0.3  &  2.3$\pm$0.3 \\
  900  & 1.5$\pm$0.1  &  1.4$\pm$0.2  &  1.9$\pm$0.2  &  3.1$\pm$0.5 \\
 1000  & 1.7$\pm$0.1  &  2.0$\pm$0.2  &  2.4$\pm$0.2  &  2.3$\pm$0.3 \\
\hline
\end{tabular}
\medskip
\caption{\it Resolution $\sigma$ as a function of the $\gamma$ beam  energy for 
the four targets. The beam encounters on its path targets from 0 to 3.}
\label{tab:sigmavsegamma}
 \end{centering}
\end{table}%
As a general trend, the $\sigma$ slightly decreases at increasing energies 
and the two deepest targets usually have larger values. The smallness of
the signal for higher energies and deeper targets together with the non 
perfect modeling of the background and high correlation between fit parameters
are responsible for departures from the general trend.
\noindent We also studied the variation of the signal as a function of 
the target thickness. 
We used a beam energy of 600~keV to ensure adeguate penetration.
Results for target thickness in the range 1--5~mm are shown in 
Fig.~\ref{fig:signalvsdz}. As expected, the signal increases as a function of 
the target thickness. Already for 1~mm thickness a signal is visible for all
the four targets although at a reduced level. 
Even if with this small change of density it would not be possible to provide
a correct image of the object, the presence of the signal would already 
fulfill the requirements needed for most NDT applications.
As an exercise we tried to obtain the target thickness using the obtained 
resolution as input to the fit function. As it can be seen in  
Table.~\ref{tab:thickness}, the capability to determine the target size 
improves with increasing target thickness and deteriorates with increasing 
depth in the object.
\begin{table}[h]
 \begin{centering}
 \begin{tabular}{crrrr}
  \hline
 \multicolumn{1}{c} {\rule{0mm}{5mm} $\Delta z$(mm)} &  
 \multicolumn{1}{r} {$\Delta z_{0}$ (mm)} & \multicolumn{1}{r} {$\Delta z_{1}$ (mm)} & 
 \multicolumn{1}{r} {$\Delta z_{2}$ (mm)} & \multicolumn{1}{r} {$\Delta z_{3}$ (mm)} \\
  \hline
  1  & 1.0$\pm$0.1  &  1.4$\pm$0.2  &  0.7$\pm$0.2  &  1.5$\pm$0.3 \\
  2  & 1.9$\pm$0.1  &  2.5$\pm$0.2  &  2.7$\pm$1.6  &  1.8$\pm$0.2 \\
  3  & 2.9$\pm$0.1  &  2.8$\pm$1.0  &  3.5$\pm$1.0  &  3.6$\pm$2.3 \\
  4  & 3.9$\pm$0.2  &  4.5$\pm$0.5  &  4.8$\pm$1.0  &  4.1$\pm$2.1 \\
  5  & 5.0$\pm$0.3  &  4.8$\pm$0.4  &  5.0$\pm$0.5  &  5.3$\pm$1.1 \\
\hline
\end{tabular}
\medskip
\caption{\it 
  Reconstructed target thickness as a function of the true thickness ($\Delta z$).
  A beam energy of 600~keV was used.}
\label{tab:thickness}
 \end{centering}
\end{table}
We also investigated the signal dependence on the target density,
Fig.~\ref{fig:signalvsdensity} shows the results. 
The signal for densities less than 1~g/cm$^3$ appears as a reversed peak
with respect to the exponentiall falling signal of the surrounding material.
The results of the simulations indicate that, at least for the first two
targets, inhomogeneities of about 6\% could be located.
Finally, in Fig.~\ref{fig:signalvsngen} we give the number of raw signal events
as a function of the number of generated events in the simulation.
\subsection{Corrections for background and attenuation}
\label{sec:attenuation}
\begin{figure}[htb]
 \begin{center}
  \includegraphics[width=9.0cm]{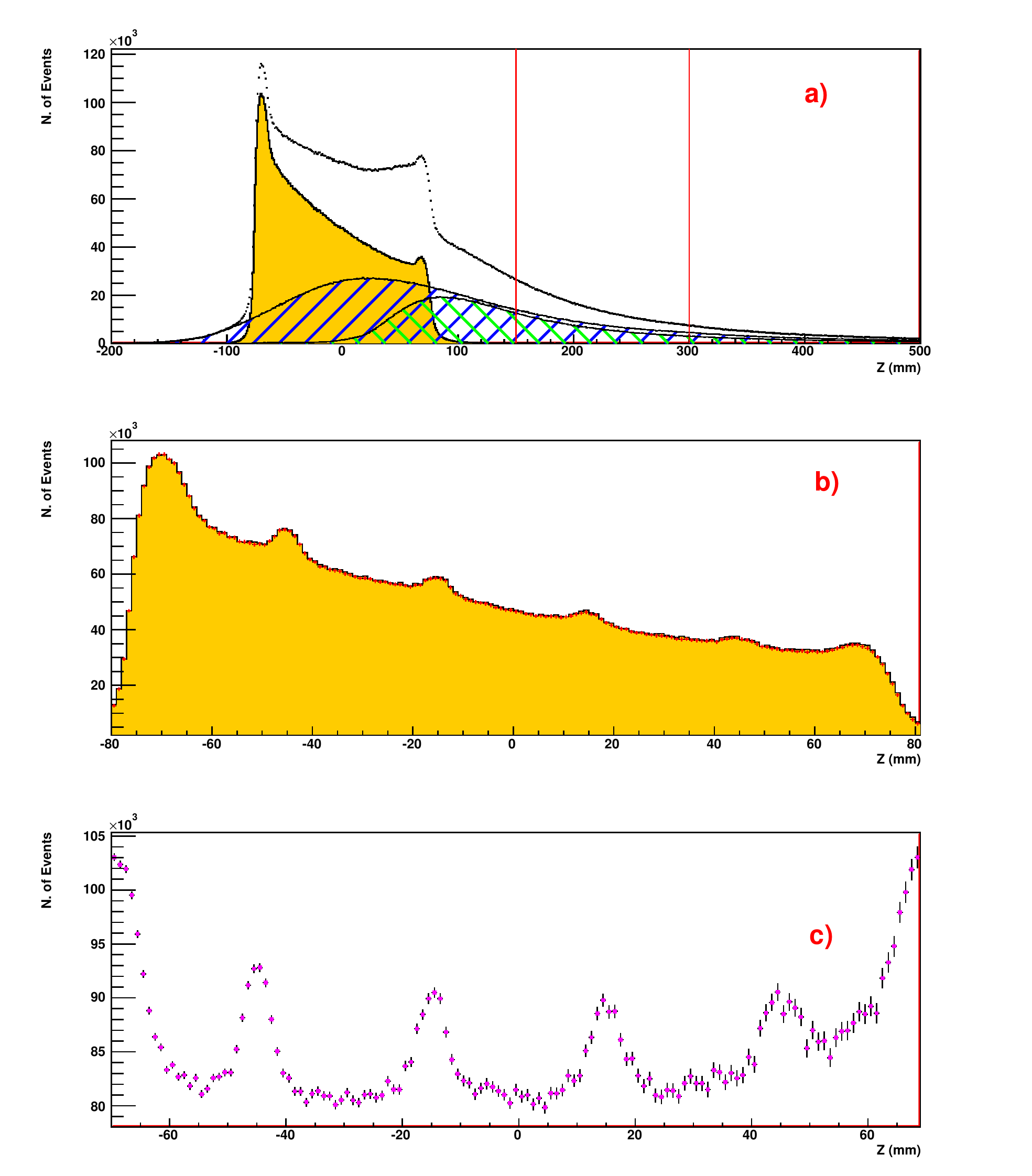}
  \caption{\it The figure shows the various steps of the procedure
   for the background correction.
    a) Reconstructed $z$ distribution  
   for the simulation without targets. This corresponds to the ''MC''
   sample of our procedure. The region between 150 and 300 mm is used
   to normalize the ''DATA'' sample.
   The shaded histogram is for S1 event type, 45$^\circ$ hatching 
   for Sn and 145$^\circ$ hatching is for Escaper type. 
   The black squares on the top represent the total distribution.
   b) $z$ distribution of ''DATA'' sample after background
   subtraction (dots) compared to the S1 event distribution
   (shaded area).
   c) $z$ distribution of ''DATA'' events after background subtraction 
   and correction for attenuation of the beam.
   A beam energy of 600 keV was used, $10^9$ events were simulated. }
  \label{fig:attencorr}
 \end{center}
\end{figure}

So far we have described the capability of the proposed device to detect
localized changes in the density of an extended object. The density
variations produce bumps (or dips) in the distribution of the reconstructed
z position. To produce an image of the interior of the object, however, 
corrections for the background induced by multiply scattered events
and Escapers, as well as for beam attenuation need to be implemented.
First we consider the background. The component induced by multiply
scattered events affects the whole z range in an almost flat way
(see Fig.~\ref{fig:attencorr}-a) while the Escaper component grows with beam
energy and affects the forward z region. The background is inherently
connected to geometrical aspects (and to the detection capabilities of
the HPGe for the Escaper component) and does not depend on the density
distribution of the object to be imaged.
We tested this feature by comparing the z distributions of multiply
scattered events and Escapers of the various simulations done with different
size and density of the targets and with targets removed. 
We also added, inside the brain volume, other volumes of different 
sizes and densities in order to simulate non uniformities in the 
density distribution. 
Finally we replaced all the head materials with water. 
All these density variations modify the scale of the background 
distribution but not the shape. 
This indicates that the MC approach can be effective in removing the unwanted 
background component also in a ''real world'' application of this device. 
We notice also that, alternatively, the background component induced
by the Escaper events could be eliminated by surrounding the HPGe
detectors with some highly sensitive scintillator, like the NaI type,
operated as a veto.\\ 
In our analysis we used the tail of the $z$ distribution in the range 
150$\div$300~mm to normalize the $z$ distribution of ''DATA'', i.e. our 
reference sample,  to ''MC'', the simulation without targets.
We chose this region because it receives contributions only from background
(see Fig.~\ref{fig:attencorr}-a).
With proper normalization we can subtract the background component and 
compare the obtained distribution to the original S1 event distribution in 
Fig.~\ref{fig:attencorr}-b.\\
To provide an indication of how a correction for the beam attenuation
could be implemented (a full treatment of this subject is beyond the scope
of this work) we consider the passage of our pencil beam through the 
object: on its path it will impinge on a small cylinder of material 
with length equal to length of the object.
We subdivide this cylinder in n equal voxels of thickness dz; 
if $N_{i}$ is the number of photons entering the $i$-th voxel 
the number of photons reaching the next voxel along z will be:
\begin{equation}
N_{i+1} = N_{i} \exp(-\mu_{i} dz) \simeq N_{i} \left( 1 - \mu_{i} dz \right)
\end{equation}
where $\mu_{i}$ is the attenuation coefficient which, given the photon energy,
is dominated by Compton scattering. In the above relation it is also 
assumed that the voxel thickness is small enough to allow the replacement of 
the exponential by its linear expansion.
The signal induced by singly scattered events in the $i-$th voxel will be:
\begin{equation}
S1_{i} = \epsilon N_{i} \mu_{i} dz
\end{equation}
where $\epsilon$ is the detection efficiency. It accounts for the probability that 
the scattered photon will not undergo to any other interaction on its way out of
the head volume and that it will produce a signal in the HPGe detector.
We assume that this efficency is the same for all the voxels, i.e. that does not 
depend on $z$. 
To correct the $i$-th signal for beam attenuation the number of $\gamma$ scattered 
in all the preceding voxels must be iteratively added. With some algebra we obtain:
\begin{equation}
S1_{i}^{corr} = S1_{i} 
\Bigl{(} 1 + \frac{\sum_{j=1}^{j=i-1}S1_{j}}{\epsilon N_{1} - \sum_{j=1}^{j=i-1}S1_{j}} \Bigr{)}
\end{equation}
where the index $j$ runs from the first voxel to the $i-1$th one and $N_{1}$ is 
the number of event impinging on the first voxel. Given the simmetry
of the set-up the first and the last voxel should give the same signal since they 
are made by the same material. This constraint can be used to 
determine the value $\epsilon N_{1}$ in the above formula.\\
The $z$ distribution, corrected for background and for attenuation is shown in
Fig.~\ref{fig:attencorr}-c. In spite of the simplicity of the outlined approach
we see a reasonable flatness of the regions corresponding to the homogeneous
brain tissue in which targets are embedded. At large $z$ however an overcorrection
appears. This is probably due to a residual $z$ dependence  of the 
detection efficiency related to the geometrical acceptance of the detector.
Also the ratio of the corrected (for attenuation and resolution effects) signal 
peaks to baseline, which should equal the ratio of the target to brain density, 
is reproduced in an acceptable way. 
A value of about 1.15 was obtained for the first three targets, to be compared to 
the true one of 1.30.
We also provide an estimate of the image contrast which could be obtained
by this device. We define the contrast C as:
\begin{equation}
C =  \frac{\bar{S}_{max} - \bar{S}_{st}}{\bar{S}_{st}}
\end{equation}
where $ \bar{S}_{max} $ is the mean of the four maxima of the signal from
the inserts and $ \bar{S}_{st} $ is the average of the regions between the 
inclusions, which correspond to the uniform soft tissue. 
A value of 12\% is obtained for a beam energy of 600 keV.
A deconvolution was performed to cross-check the results on the expected 
resolution obtained, see Table~\ref{tab:sigmavsegamma}, with data not
corrected for attenuation and background.
The fit function was a rectangular pulse convoluted with a gaussian, the true
position and thickness of the four inserts were used as input to the fit, the
background was modeled with a linear polynomial. 
The resolution $\sigma$ for the four inserts, extracted from the corrected 
data, agree with those reported in Table~\ref{tab:sigmavsegamma}.
Finally we remark that, contrary to other imaging approaches involving
Compton scattering, this method provides directly and without the need
of any backprojection technique, the density distribution of the object.
What is shown in Fig.~\ref{fig:attencorr}-c is the  image corresponding
to a portion of the object 2~mm wide in $x$ and $y$ directions (the beam size)
along its $z$ coordinate for a given beam position. The full 3D image is 
obtained by raster scanning the surface of the object and merging 
together all the acquired  images.

\section{Dose calculations}
\label{sec:dose}
\begin{figure}[htb]
 \begin{center}
  \includegraphics[width=8cm]{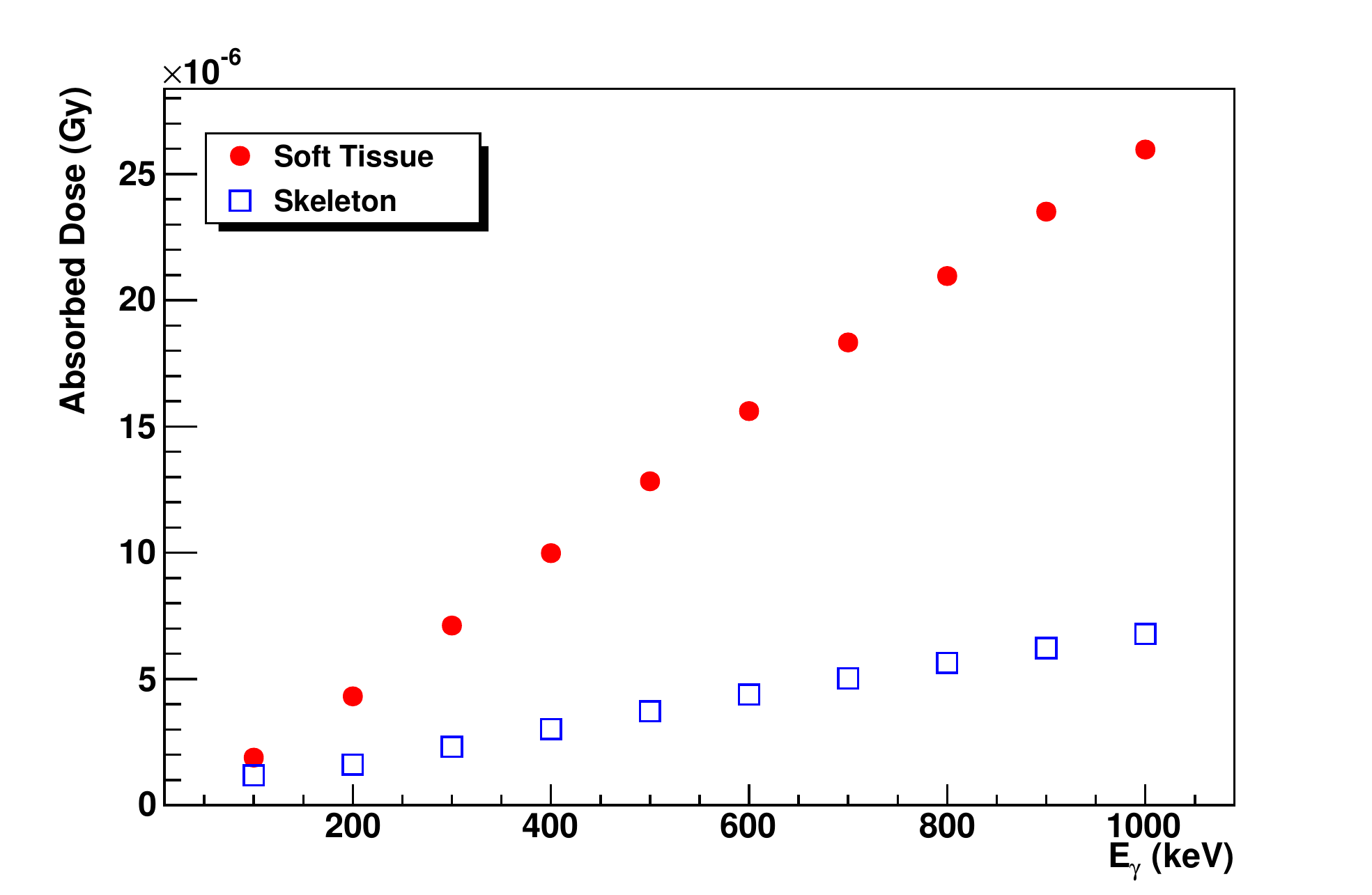}
  \caption{\it Energy released in soft (full circles) and skeleton 
   (open squares) tissue as a function of the $\gamma$ beam energy.
    For each energy $10^9$ events were generated. }
  \label{fig:edepvsegammma}
 \end{center}
\end{figure}

\begin{figure}[ht]
 \begin{center}
  \includegraphics[width=8.9cm]{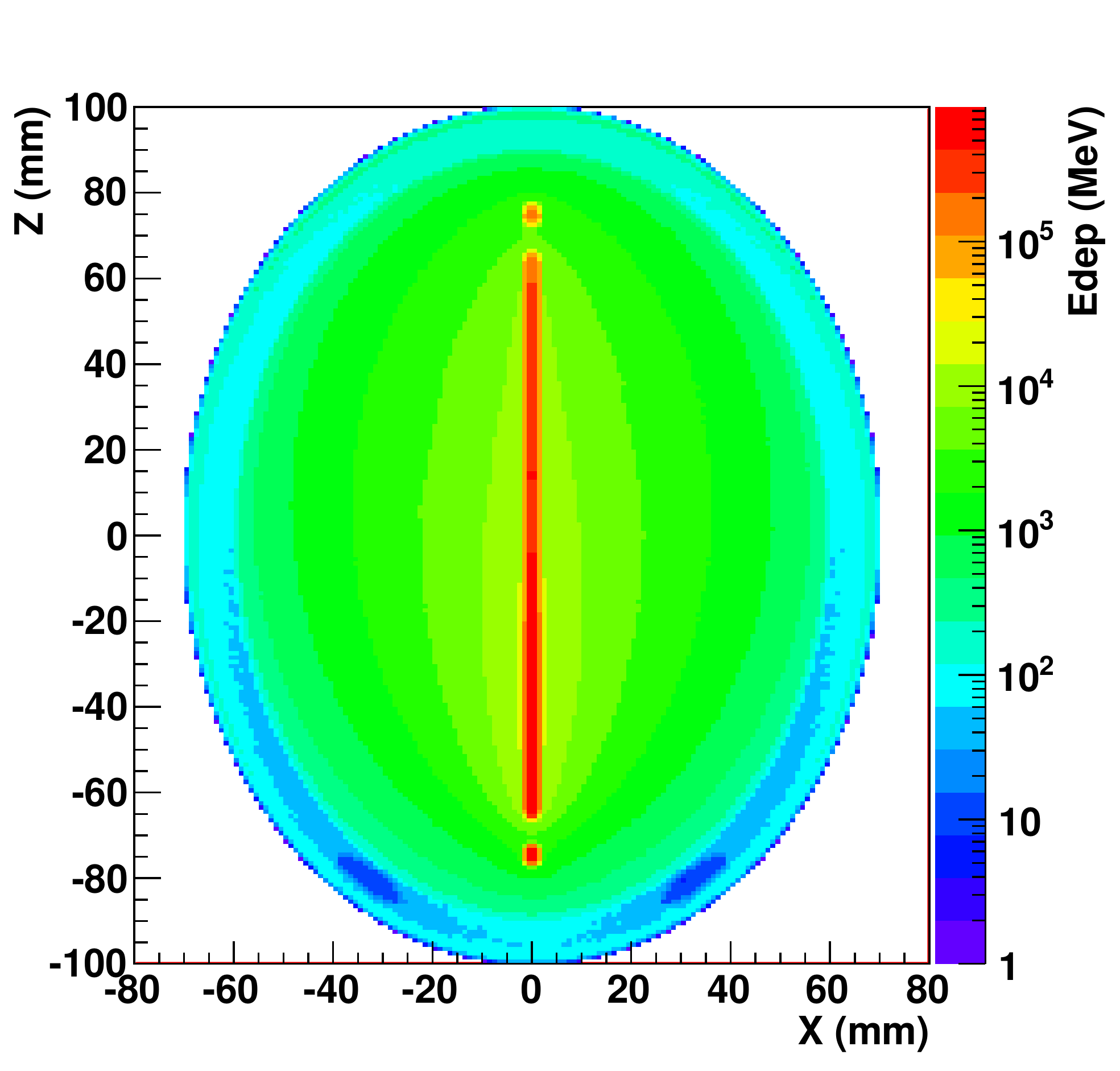}
  \caption{\it Projection onto the longitudinal plane, for one scan position, of the 
   energy deposited in soft tissue. The beam, of 600~keV energy, positioned at $x=0$, 
   is directed from negative to positive $z$. Given the collimation of the beam about 
   half of the energy deposition is concentrated along the beam path. For the simulation
   $10^9$ events were used.}
  \label{fig:dosesoft}
 \end{center}
\end{figure}

\begin{figure}[h!]
 \begin{center}
  \includegraphics[width=8.9cm]{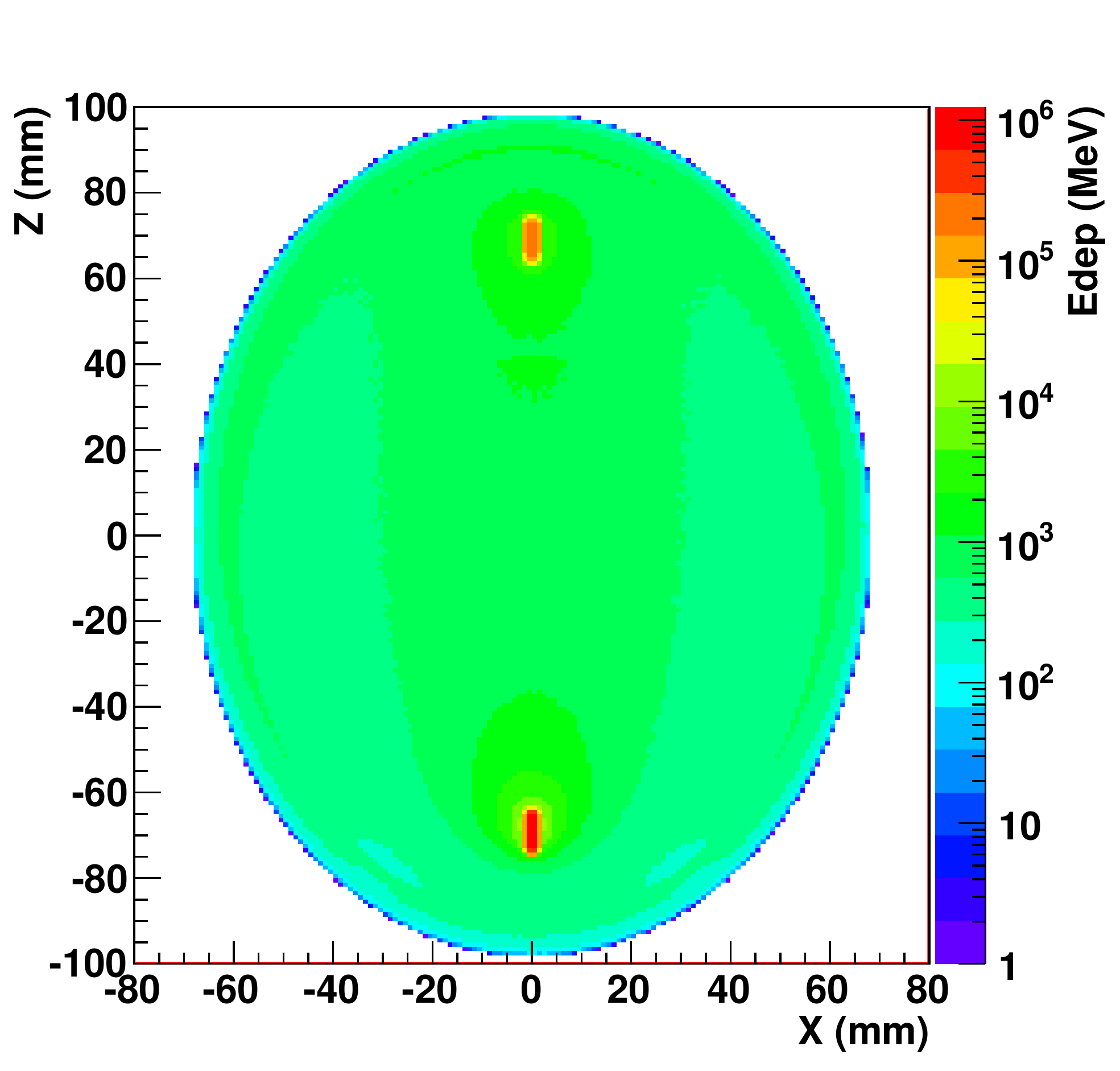}
  \caption{\it Projection onto the longitudinal plane, for one scan position, of the 
   energy deposited in skeleton tissue. The beam, of 600~keV energy, positioned at $x=0$,
   is directed from negative to positive $z$. The bottom and upper spots represent the 
   beam entrance and exit points, respectively. For the simulation $10^9$ events were used.}
  \label{fig:doseskel}
 \end{center}
\end{figure}

To calculate the absorbed dose by the soft (i.e. skin and brain) and skeleton
tissue the head volume was divided in cubic voxels of 1~mm side.
During the simulation all the particles, primary $\gamma$ 
or secondaries, were followed and step by step the energy deposited in each 
voxel, provided by the GetTotalEnergyDeposit() method of GEANT4, was stored. 
The energy released in soft and skeleton tissue, for our standard simulation
of $10^9$ events is reported as a function of the beam energy in 
Fig.~\ref{fig:edepvsegammma}. 
The energy deposited increases linearly with beam energy.
Given the highly collimated beam which is used in our application the most
of the energy deposition is concentrated in a narrow cylinder around the beam path
as can be seen in Fig.~\ref{fig:dosesoft} and Fig.~\ref{fig:doseskel}, which
show, for a beam energy of 600~keV, a longitudinal projection of the deposited 
energy in soft and skeleton tissue, respectively. 
Although almost all the voxels in which the head volume is partitioned
are interested by the release of energy, the energy deposited in 1~mm 
radius around the beam path amounts to about one half of the total.
This highly non homogeneous energy deposition makes unpracticable the usual 
way of calculating the absorbed dose i.e. dividing the total energy released 
in the voxels by their total mass. Furthermore the region to be imaged needs
to be illuminated by raster moving the $\gamma$ beam on its surface.
During a scan therefore every part of the imaged region will receive an energy
deposit like that of Figs.~\ref{fig:dosesoft} and \ref{fig:doseskel} 
(which refer to one scan position), when illuminated by the beam, and an 
additional contribution, which depends on the distance from the beam, when 
not illuminated.\\
To estimate the absorbed dose for a complete scan of the brain volume we firstly 
obtained an analytical representation of the dose as a function of the distance 
from the beam  in the transverse plane. 
We considered voxels of 2$\times$2~mm$^2$ (i.e. the beam spot size) in the 
transverse direction and 130/17~mm deep along the beam for soft/skeleton tissue, 
respectively. Then we simulated the scan procedure by considering an array of 
such voxels 80~mm wide and 40~mm high and moving the beam, in steps of 2~mm, 
upon it.  
For every beam position the dose delivered to the illuminated voxel and to all 
the other was recorded. This study indicates that the dose is almost uniformly 
distributed over the array and reaches a maximum value of 110~mGy, the contribution 
of soft and skeleton tissue being equally distribued. 
This value drops of about 15\% for voxels on the array border.
Using the radiation and tissue weighting factors recommended in~\cite{NewICRP}
we obtain for a complete scan with $10^9$ photons for each point, at an energy of 
600~keV, an effective dose of about 1~mSv which is half of the average dose imparted 
during a computed tomographic examination of the head \cite{EffDose}.

\section{Conclusions}
\label{sec:conclusions}
In this paper we have studied the potential of a novel approach of 3D 
imaging with photon scattering.\\
We studied as an application of this technique the case of brain 
imaging and show how the proposed device could be able to detect 
localized density variations inside it. Results on the attainable sensitivity 
were shown as a function of the beam energy, of the number of impinging photons
and of various target parameters: depth, thickness and density.\\
We studied in detail the signal and background characteristics and 
outlined a method, relying on the MC approach, to remove the background.
A simple model to correct the signal for beam attenuation has also been
introduced.\\
The simulation indicates that a complete scan of the head volume at 
600~keV photon energy would result in an effective dose of about 1~mSv which 
is comparable to the average dose imparted during a computed tomographic 
examination of the head.

\section{Acknowledgments}
We would like to thank Cinzia Talamonti for very helpful and stimulating discussions.

\end{document}